\documentclass[epj]{svjour}

\def\be{\begin{equation}} 
\def\ee{\end{equation}}   

\RequirePackage{graphicx}

\usepackage{cite} 
\usepackage{amsmath}
\usepackage{hyperref}
\usepackage{cite}
\usepackage{amsmath,amssymb}
\usepackage{xcolor}

\usepackage{graphics}

\usepackage{cite} 
\usepackage{amsmath}
\usepackage{hyperref}
\usepackage{cite}
\usepackage{amsmath,amssymb}
\usepackage{xcolor}

\begin{document}
\title{Greybody factors and quasinormal modes for a nonminimally coupled
scalar field in a cloud of strings in (2+1)-dimensional background}

\author{
\'Angel Rinc\'on \inst{1}
\thanks{E-mail: \href{mailto:arrincon@uc.cl}{\nolinkurl{arrincon@uc.cl}} }
\and
Grigoris Panotopoulos \inst{2} 
\thanks{E-mail: \href{mailto:grigorios.panotopoulos@tecnico.ulisboa.pt}{\nolinkurl{grigorios.panotopoulos@tecnico.ulisboa.pt}} }
}                     
%
%
\institute{ 
Instituto de F{\'i}sica, Pontificia Cat{\'o}lica Universidad de Chile,
Av. Vicu{\~n}a Mackenna 4860, Santiago, Chile
\and
Centro de Astrof\'{\i}sica e Gravita{\c c}{\~a}o, Departamento de F{\'i}sica, Instituto Superior T\'ecnico-IST,
\\
Universidade de Lisboa-UL, Av. Rovisco Pais, 1049-001 Lisboa, Portugal
}

\date{Received: date / Revised version: date}
%
\abstract{
We study the propagation of a probe massless nonminimally coupled scalar field in a fixed gravitational background of a cloud of strings in (2+1) dimensions. We obtain exact analytical expressions for the reflection coefficient, the absorption cross section, the decay rate as well as the quasinormal frequencies. The impact of the nonminimal coupling is investigated in detail. Our results show that Universality is not respected in general, and that scalar perturbations are stable.
\PACS{
      {PACS-key}{discribing text of that key}   \and
      {PACS-key}{discribing text of that key}
     } 
} 
\maketitle

\section{Introduction}\label{Intro}

Black holes (BHs hereafter), a generic prediction of Einstein's General Relativity, are objects of paramount importance both for classical and quantum gravity. Greybody factors and quasinormal modes are two topics related to black hole physics of particular interest. On the one hand, Hawking radiation \cite{hawking1,hawking2}, since it is as a manifestation of a quantum effect in curved spacetime, has always attracted a lot of interest although it has not been detected in the Universe yet. The emitted particles feel an effective potential that back scatters part of the emitted radiation back into the black hole. The greybody factor is a frequency dependent quantity that measures the deviation from the original black body radiation spectrum, and provides us with valuable information about the black hole horizon structure \cite{kanti}. The propagation and relativistic scattering of fields has been investigated both in asymptotically flat spacetimes and in background with a non-vanishing cosmological constant. For a partial list see e.g.\cite{p1,p2,massive,Fernando,p3,p4,p5,p6,p7,p8,p9,p10,coupling,p11,p12,p13,chinos,PR1,PR2,PR3} and references therein. 

On the other hand, LIGO historical direct detection of gravitational waves \cite{ligo1,ligo2,ligo3} from black hole mergers has opened a completely new window to our Universe, and allows us to test gravity and probe strong gravitational fields. Consequently, lately there is an increasing interest in black hole perturbations \cite{wheeler,zerilli1,zerilli2,zerilli3,moncrief,teukolsky} and quasinormal modes of black holes, intimately related to the ring down phase after the formation of the distorded object during the merging of two black holes. When a black hole is perturbed the geometry of spacetime undergoes dumped oscillations, which are characterized by the quasinormal modes with a non-vanishing imaginary part. Chandrasekhar'e monograph provides us with a comprehensive overview of black hole perturbations \cite{chandra}. Quasinormal modes of black holes have been extensively studied, and for excellent reviews see e.g. \cite{review1,review2,review3}.

The Ba{\~n}ados, Teitelboim and Zanelli (BTZ) black hole solution \cite{BTZ1,BTZ2,BTZ3} in (1+2) dimensions marked the birth of the interest in lower-dimensional gravity. The absence of propagating degrees of freedom as well as its deep connection to the Chern-Simons term only \cite{chern1,chern2,chern3} make three-dimensional gravity special, and at the same time a framework which allow us to get insight into realistic black holes in four dimensions by studying a mathematically simpler three-dimensional system. The BTZ black hole is sourced by a negative cosmological constant, but other possibilities, such as scalar or electromagnetic fields \cite{mann,EM1,EM2,Cataldo:2000we}, also exist. What is more, looking for a complete theory of quantum gravity, black hole solutions that admit scale-dependent couplings, have been recently investigated. For an incomplete list see \cite{Koch:2016uso,Rincon:2017ypd,Rincon:2017goj,Rincon:2017ayr,
Contreras:2017eza,Hernandez-Arboleda:2018qdo,Contreras:2018dhs,Rincon:2018lyd,Rincon:2018sgd,Rincon:2018dsq} and references therein. 

One option less studied in the literature, which leads to a black hole solution alternative to the BTZ one, is a cloud of strings \cite{letelier}. The matter contribution is described by the Nambu-Goto action, which is well-known both from string theory \cite{polchinski} and from the study of topological defects \cite{vilenkin}. For recent studies on the topic see e.g. \cite{cloud1,cloud2,cloud3,cloud4,cloud5}. The black hole solution was obtained in \cite{cloud1}, while in \cite{cloud5} the greybody factors as well as the quasinormal modes for a massless canonical scalar field in this particular background were studied. 

The relevance of a nonminimal coupling to gravity from a theoretical point of view is well-known. Quantum loop corrections will give rise to a nonminimal coupling, even if it is absent at tree level. Its role, however, in this particular context, to the best of our knowledge, has not been investigated yet. In the present article we extend the work of \cite{cloud5} regarding the propagation of a test massless scalar filed into a fixed gravitational background of a cloud of strings considering also a non-vanishing nonminimal coupling to gravity. Note, however, that here we assume Einstein's General Relativity, although in \cite{cloud5} the authors assumed a modified $f(R)$ gravitational theory.
Our work is organized as follows: In the next section we present the model and the equation for scalar perturbations, while in section three we solve the full radial equation exactly in terms of hypergeometric functions. The reflection coefficient, the greybody factor, the decay rate and the quasinormal spectrum are discussed in section four, and finally we summarize our work in the fifth section. We adopt the mostly positive metric signature $(-,+,+)$, and we work in geometrical units where the universal constants are set to unity, $\hbar=c=k_B=8 \pi G_N=1$.

\section{Gravitational background and wave equation} \label{Classical}
\noindent
We consider a model described by the action \cite{cloud1,cloud2,cloud5}
\begin{equation}
S = S_G + S_{\text{strings}}
\end{equation}
where the gravitational part $S_G$ is given by the Einstein-Hilbert term
\begin{equation}
S_G = \frac{1}{2}\int \mathrm{d}^3 x \sqrt{-g} \: R
\end{equation}
with $R$ being the Ricci scalar and $g$ being the determinant of the metric tensor $g_{\mu \nu}$,
while the second contribution due to a cloud of strings is given by
\begin{align}
S_{\text{strings}} \equiv m \int_{\Sigma} \sqrt{-h} \mathrm{d}\lambda^0 \mathrm{d}\lambda^1
\end{align}
where $\lambda^0, \lambda^1$ are the string parameters, and $h=\text{det}(h_{AB})$ with $h_{AB}$ being the string metric.
Varying with respect to $g_{\mu \nu}$ we obtain Einstein's field equations
\begin{align}
R_{\mu \nu} - \frac{1}{2} R g_{\mu \nu} &= T_{\mu \nu}
\end{align}
with $R_{\mu \nu}$ being the Ricci tensor. Adopting the coordinate system $x^\mu=(t,r,\theta)$ the energy-momentum tensor corresponding to a cloud of strings takes the form \cite{cloud2}
\begin{align}
T^{\mu}_{\nu} = \frac{\eta}{r}\text{diag}(1,1,0)
\end{align}
with $\eta$ being the coupling constant of the cloud of strings.
Assuming a static and circularly symmetric line element of the form
\begin{align}
\mathrm{d}s^2 &= -f(r) \mathrm{d}t^2 + f(r)^{-1} \mathrm{d}r^2 + r^2 \mathrm{d}\theta^2
\end{align}
the metric function $f(r)$ is found to be \cite{cloud1,cloud5}
\begin{align}
f(r) &= -M + 2 \eta r
\end{align}
with $M$ being the mass of the black hole. In order to avoid a naked singularity the coupling constant $\eta$ must be positive, and the horizon radius is computed to be $r_H = M/(2 \eta)$. The metric function may be written down equivalently as
\begin{align}
f(r) &= 2 \eta (r - r_H)
\end{align}
where we trade the mass for the horizon radius. In the rest of the article we shall 
study the propagation of a probe massless nonminimally coupled scalar field in a 
fixed gravitational background of a cloud of strings in (2+1) dimensions.

\subsection{Scalar perturbations: the wave equation} \label{Wave}
\begin{figure*}[ht]
\centering
\includegraphics[width=0.32\textwidth]{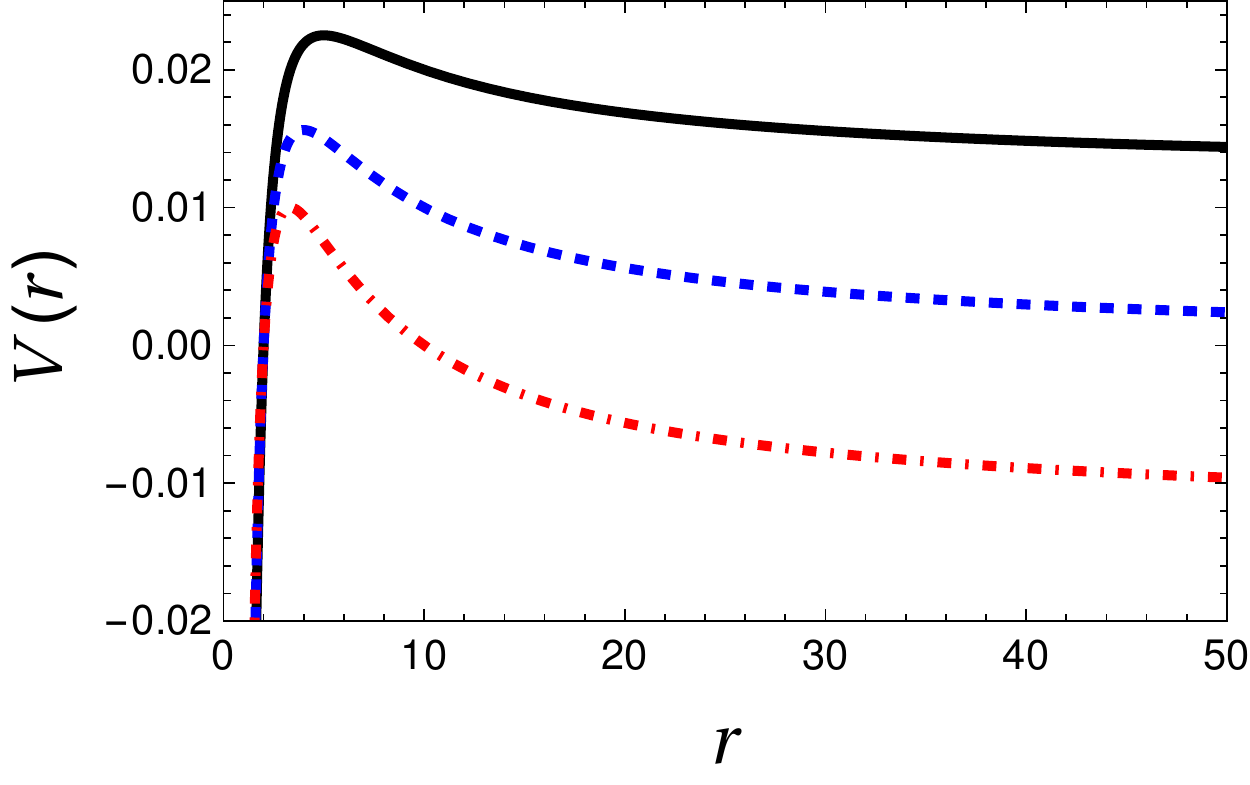}   \
\includegraphics[width=0.32\textwidth]{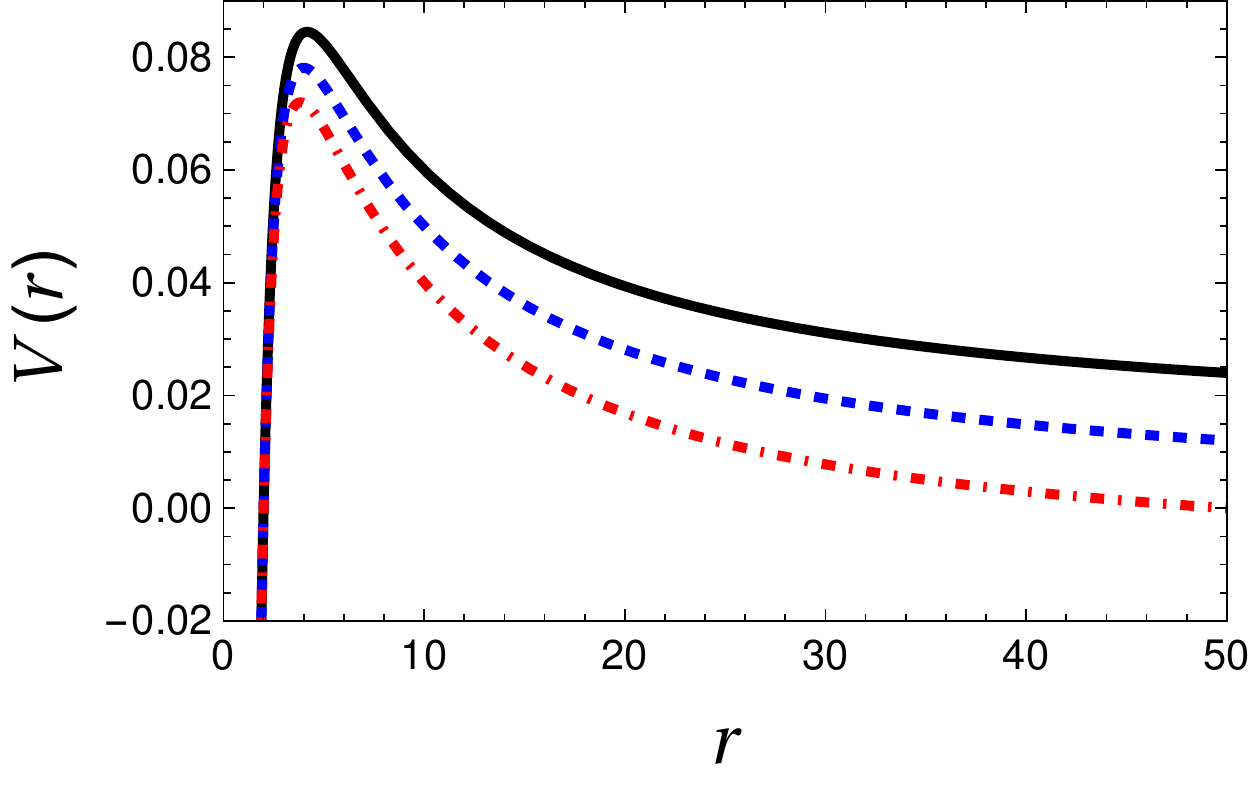}   \
\includegraphics[width=0.32\textwidth]{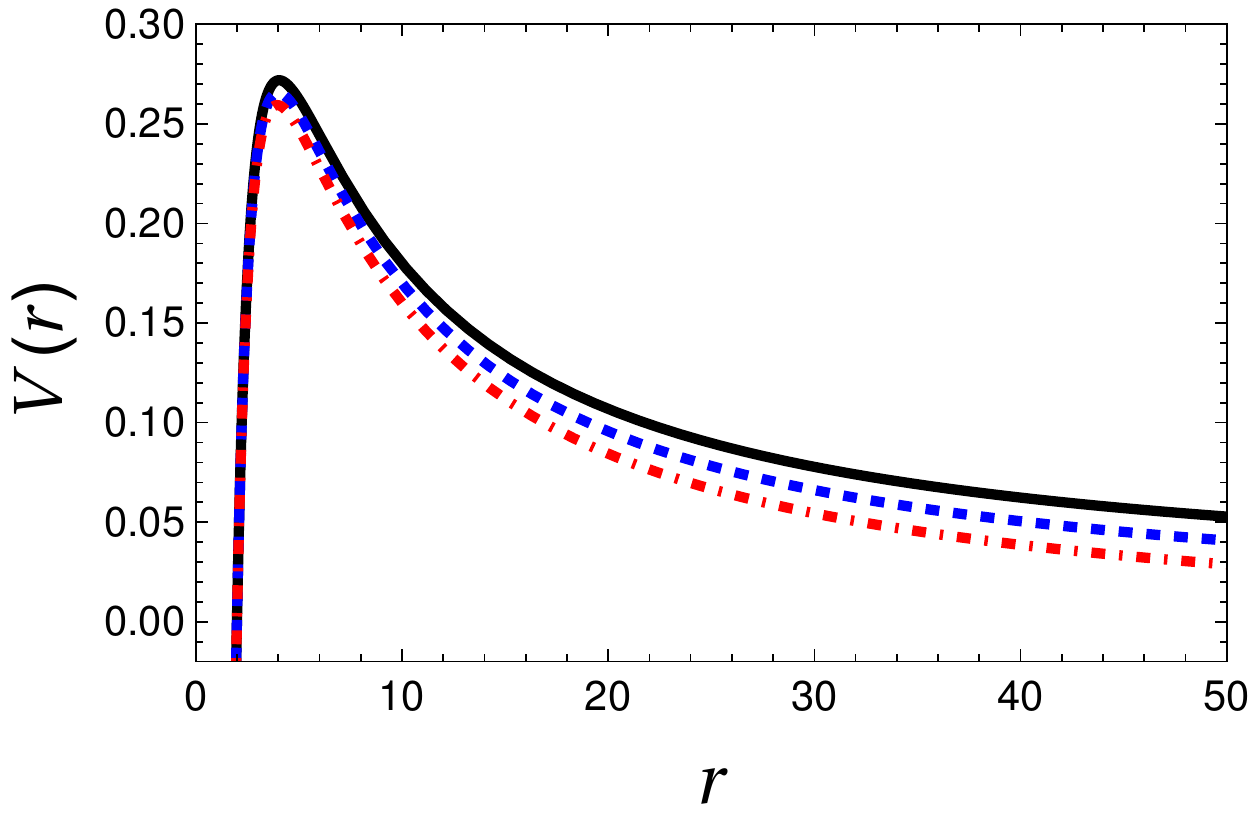}   \
\caption{
Effective potential $V(r)$ as a function of the radial coordinate $r$ assuming three different values of the parameter $m$. The panels in the first (left), second (center) and third (right) show $V(r)$ for: i) $m=0$ (case 1), ii) $m=1$ (case 2) and iii) $m=2$ (case 3), respectively. In all three figures we fix $r_H = 2$ and $\eta=0.25$. In each figure we show three different curves corresponding to: i) $\xi = 0.100$ (solid black line), ii) $\xi = 0.125$ (dashed blue line) and iii) $\xi = 0.150$ (dotted-dashed red line).
}
\label{fig:potential}
\end{figure*}

\noindent
Let us consider a probe scalar field $\Phi$ with a non-vanishing nonminimal coupling to gravity $\xi$
described by the action
\begin{equation}
S[g_{\mu \nu}, \Phi] = \frac{1}{2} \int \mathrm{d}^3x \sqrt{-g} \Bigl[  \partial^{\mu}\Phi \partial_{\mu}\Phi + \xi R_3 \Phi^2 \Bigl]
\end{equation}
The corresponding wave equation is given by the classical Klein-Gordon equation which reads
\cite{coupling,p11,p12}
\begin{equation}
\frac{1}{\sqrt{-g}} \partial_\mu (\sqrt{-g} g^{\mu \nu} \partial_\nu) \Phi = \xi R_3 \Phi
\end{equation}
where the nonminimal coupling is taken to be positive in order to maintain BH solutions (otherwise we will have naked singularity), and $R_3=-4\eta/r$ is the Ricci scalar of the cloud string background. 
Applying the usual separation of variables
\begin{equation}\label{separable}
\Phi(t,r,\phi) = e^{-i \omega t} R(r) e^{i m \phi}
\end{equation}
where $m$ is the quantum number of angular momentum, we obtain an ordinary differential equation for the radial part
\begin{equation}
R'' + \left( \frac{1}{r} + \frac{f'}{f} \right) R' + \left( \frac{\omega^2}{f^2} - \frac{m^2}{r^2 f} - \frac{\xi R_{3}}{f} \right) R = 0
\end{equation}
or rewritten it explicitly we have 
\begin{equation}
R'' + \left( \frac{1}{r} + \frac{f'}{f} \right) R' + \left( \frac{\omega^2}{f^2} - \frac{m^2}{r^2 f} + \frac{4 \eta \xi}{r f} \right) R = 0
\end{equation}
In order to read-off the effective potential barrier that the probe scalar field feels, we define new variables as follows
\begin{eqnarray}
R & = & \frac{\psi}{\sqrt{r}} \\
x & = & \int \frac{\mathrm{d}r}{f(r)}
\end{eqnarray}
where $x$ is the so-called tortoise coordinate given by 
\begin{equation}
x = \frac{1}{2 \eta} \ln (r-r_H)
\end{equation}
and recast the equation for the radial part into a Schr{\"o}dinger-like equation of the form
\begin{equation}
\frac{\mathrm{d}^2 \psi}{\mathrm{d}x^2} + (\omega^2 - V(x)) \psi = 0
\end{equation}
Therefore we obtain for the effective potential barrier the expression
\begin{equation}
V(r) = f(r) \: \left( - \frac{4 \eta \xi}{r} + \frac{m^2}{r^2}+\frac{f'(r)}{2 r}-\frac{f(r)}{4 r^2} \right)
\end{equation}
The effective potential as a function of the radial distance can be seen in Fig. \eqref{fig:potential}
for three different values of the coupling $\xi$.
The effective potential vanishes at $r=r_H$, it reaches a maximum value $V_{\text{max}}$, and tends to a constant $V_0$ when $r \rightarrow \infty$
\begin{eqnarray}
r_{\text{max}} & = & \frac{2 m^2 + r_H \eta}{m^2+4 \xi r_H \eta} \: r_H \\
V_{\text{max}} & = & \frac{\eta}{r_H} \: \frac{m^2 + (1-4 \xi) r_H \eta}{2 m^2 + r_H \eta} \\
V(r) & \rightarrow & \eta^2 (1-8 \xi) \equiv V_0, \; \; \; \; \; \; r \rightarrow \infty
\end{eqnarray}
Since the effective barrier potential vanishes at the horizon, close to the horizon $\omega^2 \gg V(x)$, and the solution for the
Schr{\"o}dinger-like equation is given by
\begin{equation}
\psi(x) = A_- e^{-i \omega x} + A_+ e^{i \omega x}
\end{equation}
Requiring purely ingoing solution \cite{kanti, Fernando, chinos} we set 
$A_+=0$ in the following. Furthermore, when $r \rightarrow \infty$ the wave equation admits plane wave solutions of the form $\psi \sim e^{\pm i \Omega x}$ when $\omega^2-V_0 > 0$, with $\Omega=\sqrt{\omega^2-V_0}$ being the modified frequency.

\section{Solution of the radial equation} \label{Radial}

Given the metric function, we observe that in the far field region, $r \gg r_H$, it takes the simple form $f(r) \sim 2 \eta r$. In addition, the angular momentum term is subdominant compared to the other two terms, and thus the differential equation for the radial part takes the form
\begin{equation}
r^2 R''(r) + 2 r R'(r) + k R(r) = 0 
\end{equation}
where the parameter $k$ is defined according to
\begin{align}
k \equiv \frac{1}{4}\left( \frac{\omega^2}{\eta ^2} + 8 \xi \right)
\end{align}
Thus, the  algebraic equation (which is Euler's equation) admits power-law solutions of the form $R(r) \sim r^\rho$. The algebraic equation for $\rho$ satisfies 
\begin{equation}
\rho ^2 + \rho + k = 0
\end{equation}
The roots of the above equation are given by
\begin{equation}
\rho_{\pm} = \frac{1}{2} \left(-1 \pm \sqrt{1-4 k}\right)
\end{equation}
which are real when $1-4 k \geq 0$ and complex when $4 k > 1$. In the following we shall consider the case of complex roots, and therefore the far field solution is given by
\begin{equation}
R_{FF}=D_- \left( \frac{r}{r_H} \right)^{\rho_-} + D_+ \left( \frac{r}{r_H} \right)^{\rho_+}
\end{equation}
where $D_-,D_+$ are two arbitrary coefficients, while the roots are given by
\begin{equation}
\rho_{\pm} = \frac{1}{2} 
\Bigg( -1 \pm i \sqrt{\frac{\omega^2}{\eta ^2} + 8 \xi -1} \Bigg)
\end{equation}
\noindent
To find the solution of the full radial equation we introduce the dimensionless parameter $z$ which is defined as follow
\begin{align}
z=1-\frac{r_H}{r}
\end{align}
and it takes values between 0 and 1. Then the new differential equation with respect to $z$ becomes
\begin{equation}
z (1-z) R_{zz} + (1-z) R_z \ +
\left( \frac{A}{z} + \frac{B}{-1+z} - C \right) R = 0
\end{equation}
where the three constants are given by
\begin{eqnarray}
A & = & \frac{\omega^2}{4 \eta^2} \\
B & = & -A - 2\xi \\
C & = & \frac{m^2}{2 \eta r_H}
\end{eqnarray}
The last differential equation can be recast in the form of the Gauss' hypergeometric equation by
removing the poles in the last term through the ansatz
\begin{equation}
R = z^\alpha (1-z)^\beta F
\end{equation}
where now $F$ satisfies the following differential equation
\begin{align}
\begin{split}
z (1-z) F_{zz} + &[1+2 \alpha - (1+2 \alpha+2 \beta) z] F_z +
\\
&\left( \frac{\bar{A}}{z} + \frac{\bar{B}}{-1+z} - \bar{C} \right) F = 0
\end{split}
\end{align}
and the new constants are given by
\begin{eqnarray}
\bar{A} & = & A + \alpha^2 \\
\bar{B} & = & B + \beta - \beta^2 \\
\bar{C} & = & C+(\alpha+\beta)^2
\end{eqnarray}
Demanding that $\bar{A}=0=\bar{B}$, we determine the parameters $\alpha$ and $\beta$ as follows
\begin{eqnarray}
\alpha & = & \pm i \frac{\omega}{2 \eta} \\
\beta & = &\frac{1}{2} \left(1 \pm i \sqrt{-1 + \frac{\omega^2}{\eta ^2} + 8 \xi }\right)
\end{eqnarray}
and finally we obtain the hypergeometric equation
\begin{equation}
z (1-z) F_{zz} + [c-(1+a+b) z] F_z - ab F = 0
\end{equation}
with parameters $a,b,c$ given by
\begin{eqnarray}
c & = & 1 + 2 \alpha \\
a & = &  \alpha + \beta + i \sqrt{C} \\
b & = & \alpha + \beta - i \sqrt{C}
\end{eqnarray}
Note that the parameters $a,b,c$ satisfy the condition $c-a-b=1-2 \beta$.
Therefore the general solution for the radial part is given by \cite{chinos}
\begin{equation}
\begin{split}
R(z) = & z^\alpha (1-z)^\beta \Bigl[ C_1 F(a,b;c;z) \ + 
\\
& C_2 z^{1-c} F(a-c+1,b-c+1;2-c;z) \Bigl]
\end{split}
\end{equation}
where $C_1,C_2$ are two arbitrary coefficients,
and the hypergeometric function can be expanded in a Taylor series as follows \cite{handbook}
\begin{equation}
F(a,b;c;z) = 1 + \frac{a b}{c} \: z+ \cdots
\end{equation}
Setting $C_2=0$ and for the choice for $\alpha= - i \omega /(2 \eta)$ we recover the purely ingoing solution close to the horizon, $R \sim (r-r_H)^\alpha$, or $\psi \sim e^{-i \omega x}$.
Therefore in the following we consider the first solution only, namely
\begin{equation}
R(z) = D z^\alpha (1-z)^\beta F(a,b;c;z)
\end{equation}
where now we have replaced $C_1$ by $D$. The sign in the expression for $\beta$ does not really matter, and in the following we consider the plus sign.

In order to match with the far field solution obtained earlier (where now $z \rightarrow 1$) we use the transformation \cite{handbook}
\begin{equation}
\begin{split}
F(a,b;c;z) = \ &\frac{\Gamma(c) \Gamma(c-a-b)}{\Gamma(c-a) \Gamma(c-b)} \times \:
\\
&F(a,b;a+b-c+1;1-z) \ +
 \\
(1-z)^{c-a-b} &\frac{\Gamma(c) \Gamma(a+b-c)}{\Gamma(a) \Gamma(b)} \times \:
\\
&F(c-a,c-b;c-a-b+1;1-z)
\end{split}
\end{equation}
and therefore the radial part as $z \rightarrow 1$ reads
\begin{equation}
\begin{split}
R \simeq  \frac{D (1-z)^\beta \Gamma(1+2 \alpha) \Gamma(1-2 \beta)}{\Gamma(1+\alpha-\beta-i \sqrt{C}) \Gamma(1+\alpha-\beta+i \sqrt{C})}
\\
+ \frac{ D (1-z)^{1-\beta} \Gamma(1+2 \alpha) \Gamma(-1+2 \beta)}{\Gamma(\alpha+\beta-i \sqrt{C}) \Gamma(\alpha+\beta+i \sqrt{C})}
\end{split}
\end{equation}
Note that $-\beta = \rho_{-}$ and $\beta-1=\rho_{+}$, and since $z=1-(r_H/r)$ the radial part $R(r)$ for $r \gg r_H$ can be written down as follows
\begin{equation}
R \simeq  D_- \left(\frac{r}{r_H}\right)^{\rho_{-}} + D_+ \left(\frac{r}{r_H}\right)^{\rho_{+}}
\end{equation}
where we have introduced $D_-, D_+$ in terms of $D$ as follows
\begin{align}
D_{-} = & \ \frac{D \Gamma(1+2 \alpha) \Gamma(1-2 \beta)}{\Gamma(1+\alpha-\beta-i \sqrt{C}) \Gamma(1+\alpha-\beta+i \sqrt{C})} \\
D_{+} = & \ \frac{D \Gamma(1+2 \alpha) \Gamma(-1+2 \beta)}{\Gamma(\alpha+\beta-i \sqrt{C}) \Gamma(\alpha+\beta+i \sqrt{C})}
\end{align}

\section{Greybody factors and quasinormal modes}

\subsection{Absorption cross section and decay rate}

\begin{figure*}[ht]
\centering
\includegraphics[width=0.48\textwidth]{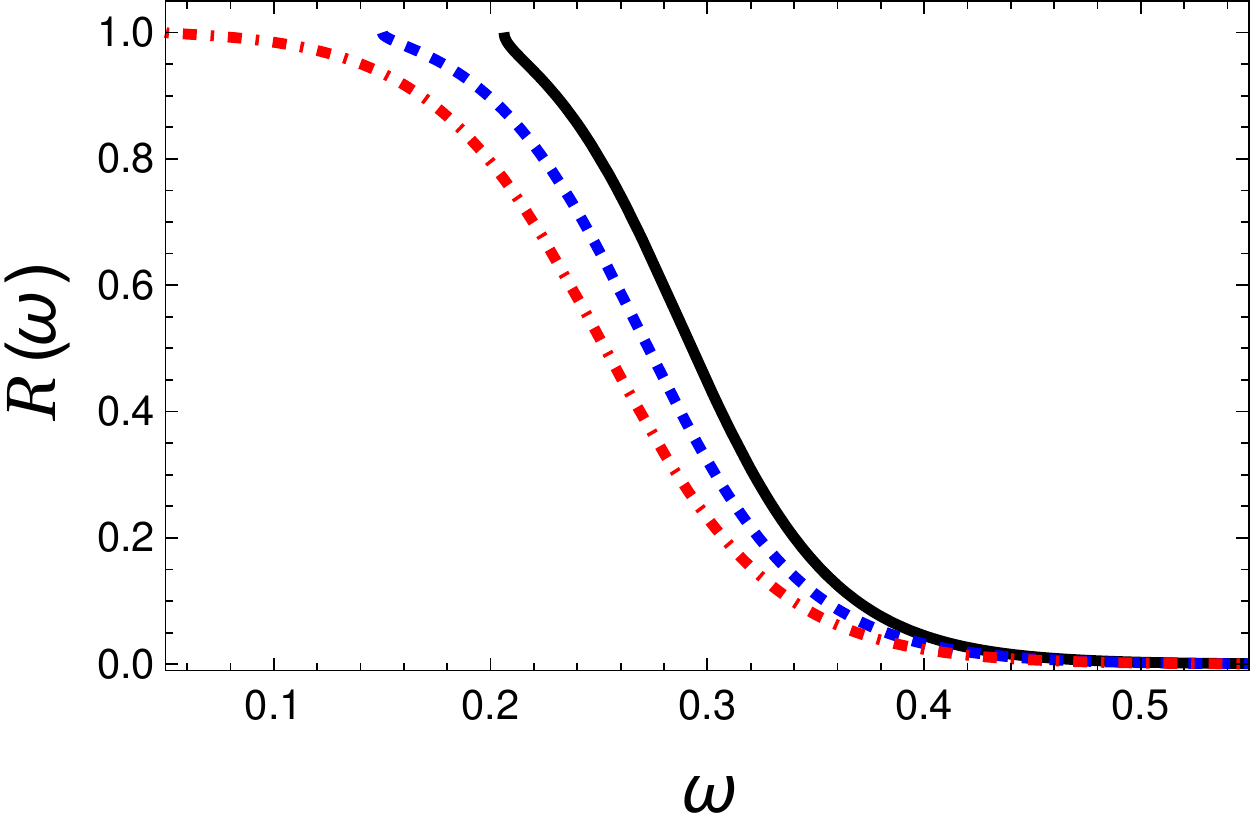}   
\ \ \
\includegraphics[width=0.48\textwidth]{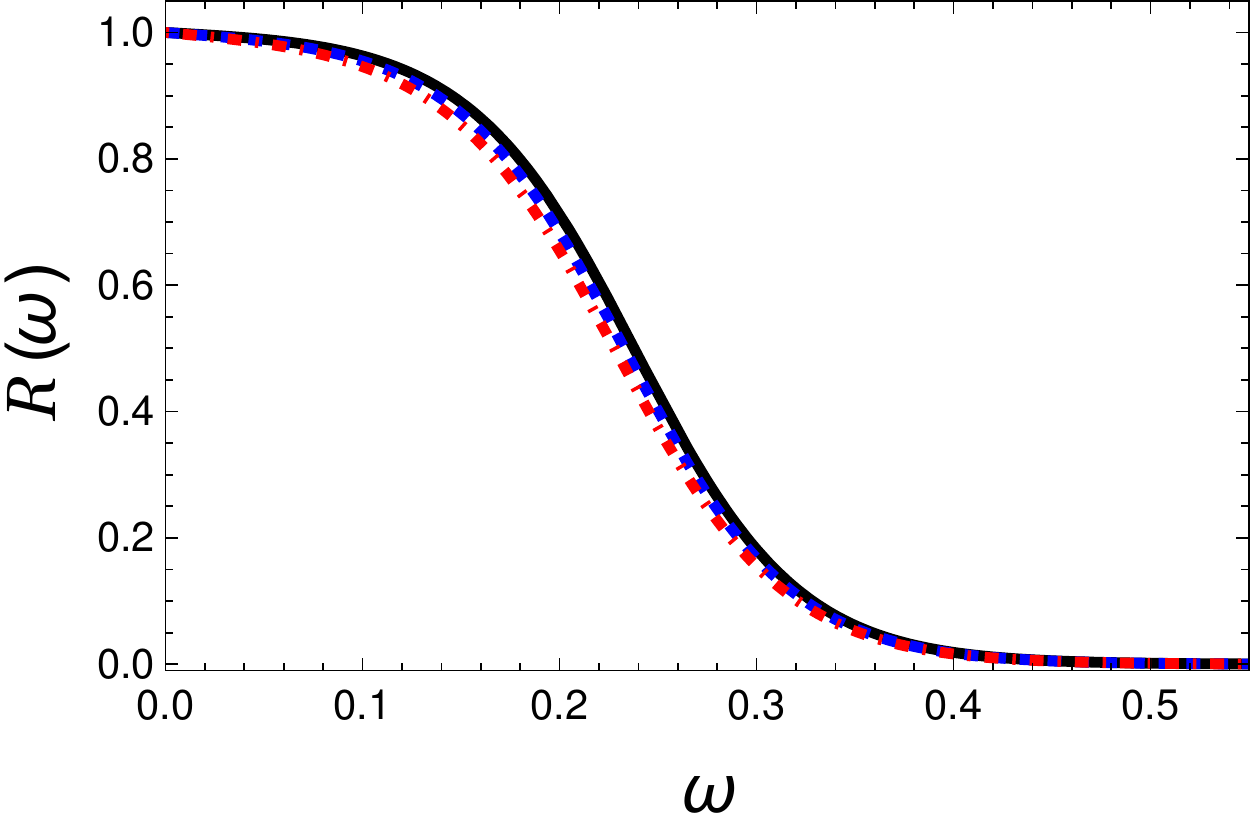}   

\caption{
Reflection coefficient $\mathcal{R}$ as a function of $\omega$ for two different regimes regarding the non-minimal coupling: i) when $ 0 \leq \xi < \xi_c$ (left) and ii) when $\xi \geq \xi_c$ (right). {\bf{Left panel:}} Reflection coefficient versus frequency for $\xi=0.04$ (solid black line), $\xi=0.08$ (dashed blue line), and $\xi=0.12$ (dotted-dashed red line).  {\bf{Right panel:}} Reflection coefficient versus frequency for $\xi=0.15$ (solid black line), $\xi=0.16$ (dashed blue line) and, $\xi=0.17$ (dotted-dashed red line). In both figures we fix $m = 1$, $r_H=2$ and $\eta = 0.25$.
}
\label{fig:Reflection}
\end{figure*}

\noindent
First we compute the reflection coefficient defined by
\begin{equation}
\mathcal{R} = \left| \frac{\textrm{outgoing wave}}{\textrm{ingoing wave}} \right|^2
\end{equation}
and therefore it is computed by $\mathcal{R} = |D_+/D_-|^2$.  Using the following identities
for the $\Gamma$ function \cite{Fernando}
\begin{eqnarray}
|\Gamma(i y)|^2 & = & \frac{\pi}{y \sinh(\pi y)}  \\
\bigg|\Gamma\left(\frac{1}{2}+i y\right)\bigg|^2 & = & \frac{\pi}{\cosh(\pi y)}
\end{eqnarray}
we obtain the final  expression
\begin{equation}\label{R1}
\mathcal{R} = \frac{\text{cosh}(\pi y_1) \text{cosh}(\pi y_2)}{\text{cosh}(\pi y_3) \text{cosh}(\pi y_4)}
\end{equation}
where $y_i$ are given by the following expressions
\begin{eqnarray}
y_1 & = & \frac{\sqrt{(\omega/\eta)^2+8 \xi-1}}{2} - \frac{\omega}{2 \eta} + \frac{|m|}{\sqrt{2 r_H \eta}} \\
y_2 & = & \frac{\sqrt{(\omega/\eta)^2+8 \xi-1}}{2} - \frac{\omega}{2 \eta} - \frac{|m|}{\sqrt{2 r_H \eta}} \\
y_3 & = & \frac{\sqrt{(\omega/\eta)^2+8 \xi-1}}{2} + \frac{\omega}{2 \eta} + \frac{|m|}{\sqrt{2 r_H \eta}} \\
y_4 & = & \frac{\sqrt{(\omega/\eta)^2+8 \xi-1}}{2} + \frac{\omega}{2 \eta} - \frac{|m|}{\sqrt{2 r_H \eta}} 
\end{eqnarray}
We see that one may consider two sectors in the problem separated by the critical value $\xi_c \equiv 1/8$. Therefore,
in the rest of the discussion we shall consider two separate cases, namely the case of weak nonminimal coupling, $0 \leq \xi < \xi_c$, and strong coupling, $\xi \geq \xi_c$. 
The reflection coefficient as a function of the frequency can be seen in Fig. \eqref{fig:Reflection} for weak (left) and strong (right) nonminimal coupling, respectively. Notice that in the weak regime there is for $\omega$ a minimum allowed value that depend on the nonminimal coupling, $\omega_{\text{min}} = 2 \sqrt{2} \eta \sqrt{\xi_c - \xi}$. The reflection coefficient is a monotonically deceasing function of the frequency, it starts at $R_{\text{ini}}=1$ and eventually it tends to zero. Furthermore, we see that when $\xi$ increases the curves are shifted downwards.

\begin{figure*}[ht]
\centering
\includegraphics[width=0.48\textwidth]{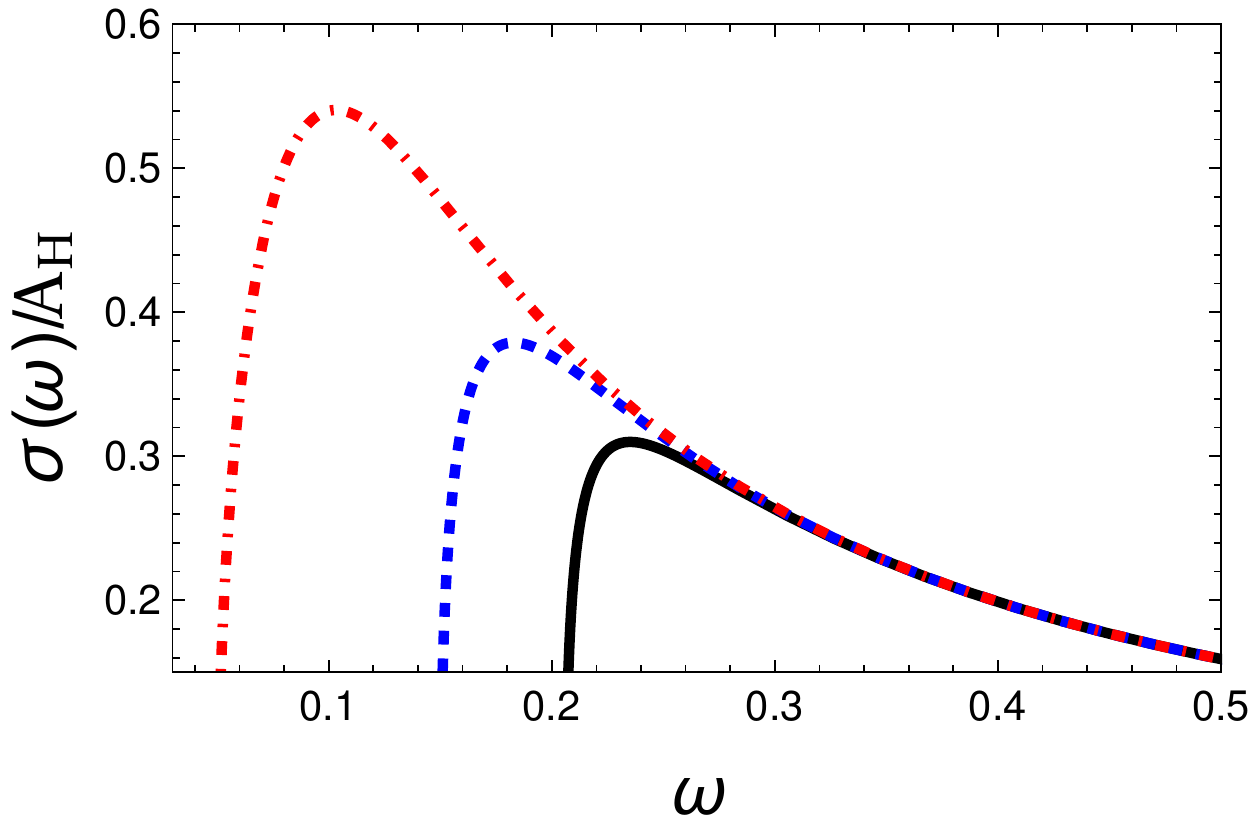}   
\ \ \
\includegraphics[width=0.48\textwidth]{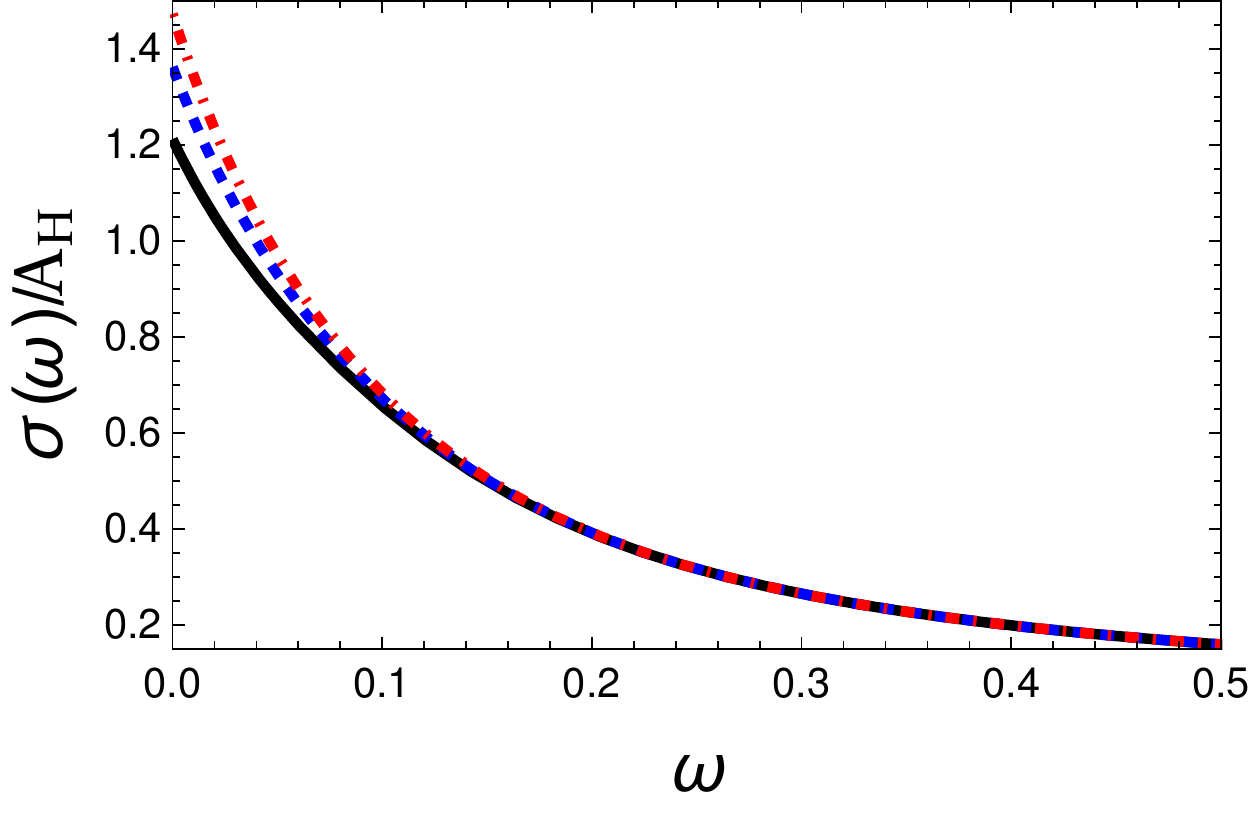}   

\medskip

\includegraphics[width=0.48\textwidth]{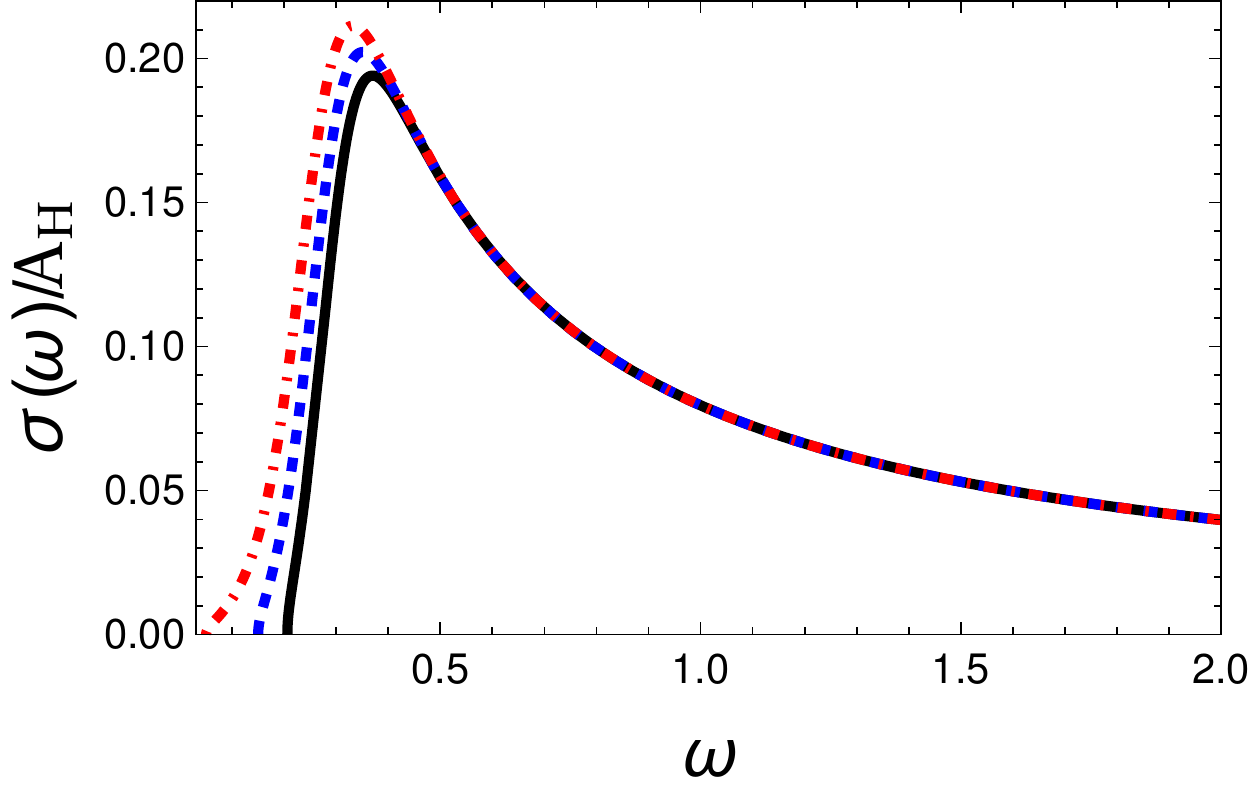}   
\ \ \
\includegraphics[width=0.48\textwidth]{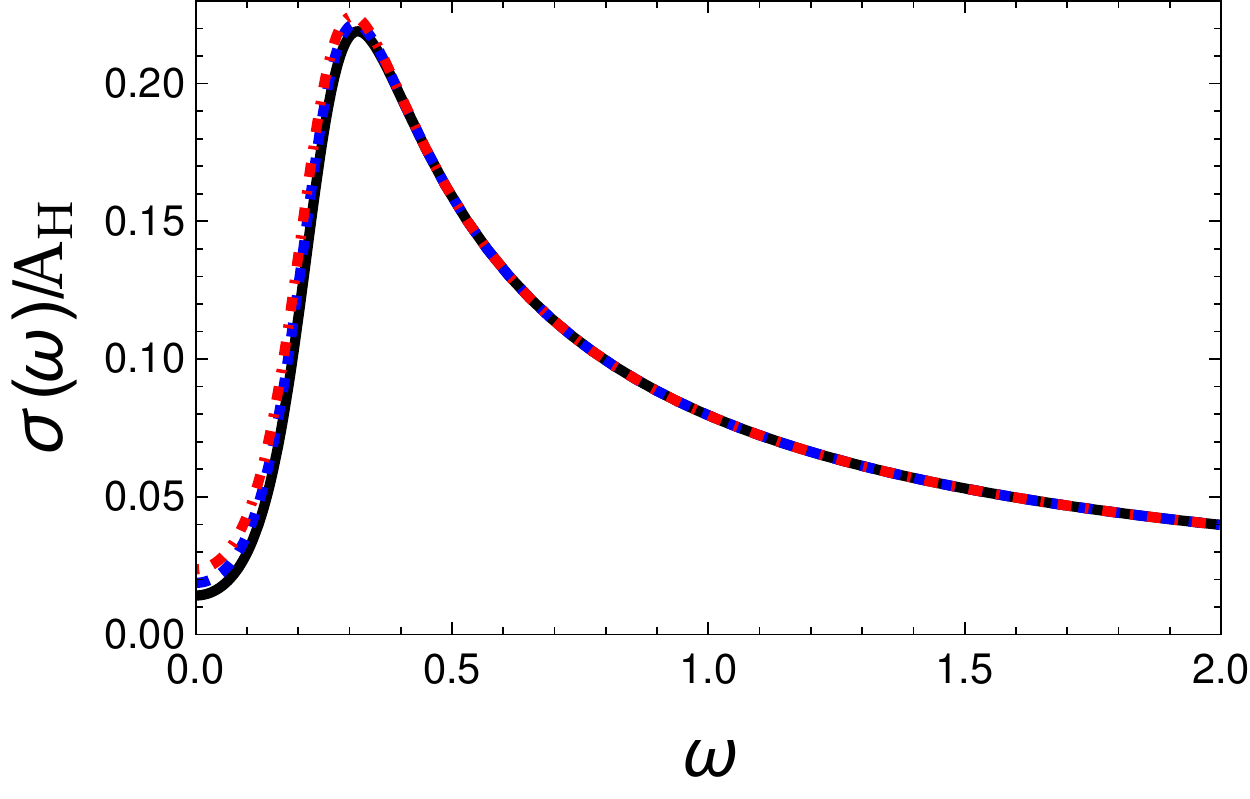} 

\caption{
$\sigma_{\text{abs}}/\mathcal{A}_H$ as a function of $\omega$ for two different regimes regarding the non-minimal coupling: i) when $ 0 \leq \xi < \xi_c$ (left) and ii) when $\xi \geq \xi_c$ (right). {\bf{Left panel:}} $\sigma_{\text{abs}}/\mathcal{A}_H$ versus frequency for $\xi=0.04$ (solid black line), $\xi=0.08$ (dashed blue line), and $\xi=0.12$ (dotted-dashed red line). {\bf{Right panel:}} $\sigma_{\text{abs}}/\mathcal{A}_H$ versus frequency for $\xi=0.15$ (solid black line), $\xi=0.16$ (dashed blue line), and $\xi=0.17$ (dotted-dashed red line). In all four figures we fix $r_H=2$ and $\eta = 0.25$. In the first row (first two figures) we assume $m=0$, while in the second row (last two figures) we assume $m=1$.
}
\label{fig:SigmaYArea}
\end{figure*}

Then, the absorption cross section is given by the simple formula \cite{Fernando,universality}
\begin{equation}
\sigma_{\text{abs}}(\omega) = \frac{1-\mathcal{R}(\omega)}{\omega}
\end{equation}
which is the three-dimensional version of the optical theorem \cite{kanti}.
In \cite{universality} it was shown that for a generic spherically symmetric black hole the absorption cross section of a minimally coupled massless scalar field for vanishing angular momentum in the low energy regime tends to a constant that coincides with the area of the horizon. It is not obvious that this still holds for a massive or a nonminimally coupled scalar field. In fact in \cite{massive} it was found that Universality was respected under certain conditions, while in \cite{coupling} it was shown that for a four-dimensional nonminimally coupled scalar field the absorption cross-section in the low energy regime tends to zero like $\omega^2$. In this work we find that as a function of $\omega$ for $m=0$ the absorption cross-section starts from a constant and eventually goes to zero, but this constant does not necessarily coincide with the area of the horizon $\mathcal{A}_H=2 \pi r_H$. We found very similar results in a previous work of ours \cite{PR1}. We define the dimensionless parameter $\sigma_{\text{abs}}/\mathcal{A}_H$ and we plot it
as a function of the frequency in Fig. \eqref{fig:SigmaYArea}. The constant increases with the coupling and finally acquires a limiting value when the coupling becomes sufficiently large. 

In the weak coupling regime as well as in the strong coupling regime, for $m > 0$, the greybody factor reaches a maximum value, and then it tends to zero decreasing monotonically. What is more, the maximum value increases with $\xi$, which shifts the curves upwards. In the strong coupling regime for $m=0$, however, the greybody factor is a monotonically decreasing function tending to zero starting from its maximum value at the origin. Similarly to the weak coupling regime, when  $\xi$ increases the curves are shifted upwards.

Since the spectrum emitted by the black hole (in three spatial dimensions) is given by \cite{coupling}
\begin{equation}\label{flux}
\frac{\mathrm{d}N(\omega)}{\mathrm{d}t} = \sum_{\ell} \frac{\sigma_{\ell}(\omega)}{e^{\omega/T_H} - 1} \frac{\mathrm{d}^3 k}{(2 \pi)^3}
\end{equation}
we define the decay rate of the black hole $\Gamma_{\text{decay}}$ by \cite{Fernando}:
\begin{align}
\Gamma_{\text{decay}}(\omega) = \frac{\sigma_{\text{abs}}(\omega)}{e^{\omega/T_H} - 1}
\end{align}
where the Hawking temperature of the cloud string black hole is given by $T_H=\eta/2\pi$. 

\begin{figure*}[ht]
\centering
\includegraphics[width=0.48\textwidth]{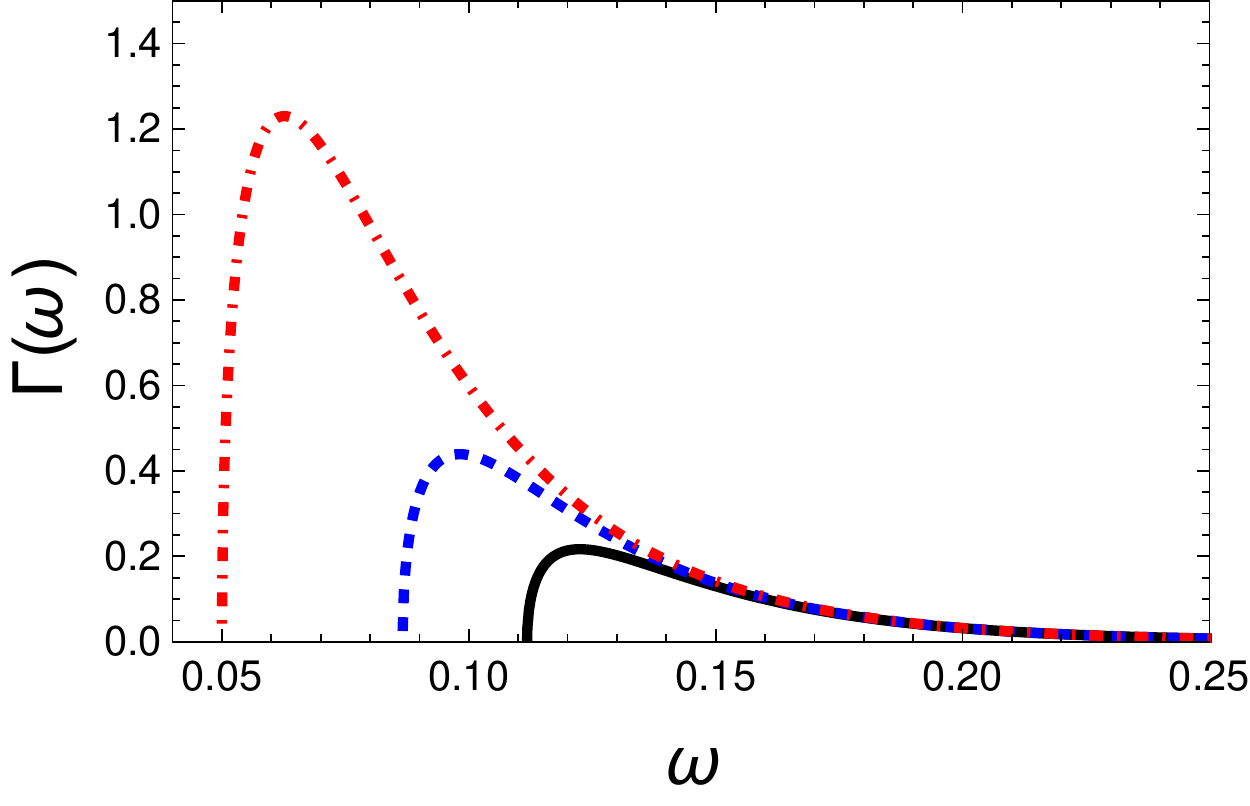}   
\ \ \
\includegraphics[width=0.48\textwidth]{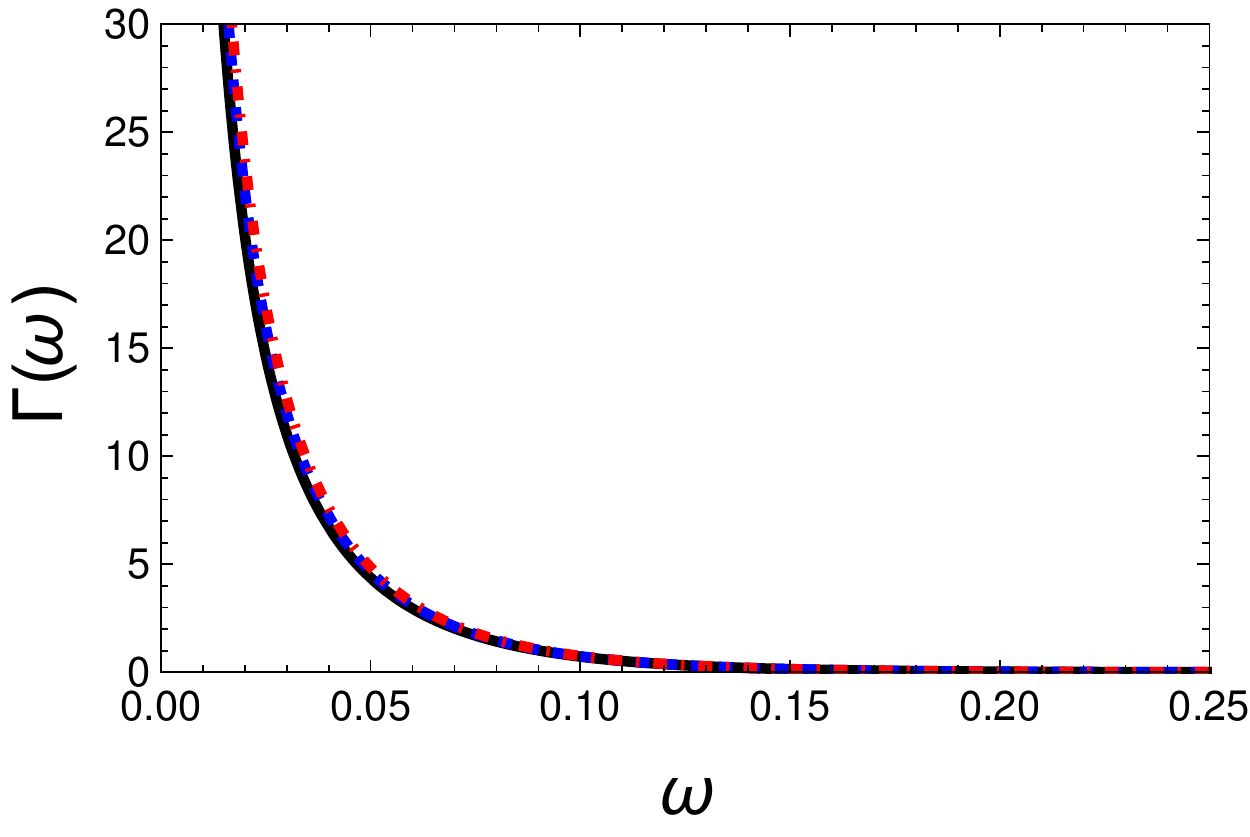}   

\caption{
Decay rate $\Gamma_{\text{decay}}$ as a function of $\omega$ for two different regimes, weak and strong nonminimal coupling: i) when $ 0 \leq \xi < \xi_c$ (left) and ii) when $\xi \geq \xi_c$ (right). {\bf{Left panel:}} $\Gamma_{\text{decay}}$ versus frequency for $\xi=0.10$ (solid black line), $\xi=0.11$ (dashed blue line), and $\xi=0.12$ (dotted-dashed red line).  {\bf{Right panel:}} $\Gamma_{\text{decay}}$ versus frequency for $\xi=0.15$ (solid black line), $\xi=0.16$ (dashed blue line), and $\xi=0.17$ (dotted-dashed red line). In both figures we fix $m = 1$, $r_H=2$ and $\eta = 0.25$.
}
\label{fig:Decay}
\end{figure*}

The decay rate as a function of frequency can be seen in the Fig. \eqref{fig:Decay} for low and large nonminimal coupling (left and right figures), respectively. Our figures are consistent with those of \cite{Fernando} in the high energy regime: the curves asymptotically go to zero. As before, when $\xi$ increases the curves are shifted upwards. In the weak coupling regime the decay rate reaches a maximum value, while in the strong coupling regime the decay rate is a monotonically decreasing function going eventually to zero.

\subsection{Quasinormal spectrum}

First we remark that computing the QNMs of black holes analytically is possible only in some cases, see e.g. \cite{potential,ferrari,cardoso2,exact1,exact2,exact3,exact4,exact5,roman1,exact6,Ovgun:2018gwt}, see also \cite{roman2} for the influence of the back reaction of the Hawking radiation upon black hole quasinormal modes. To obtain the quasinormal modes we apply the quasinormal boundary condition, according to which at infinity purely outgoing solution is required. Therefore we require that $D_-=0$ which is satisfied when the Gamma functions in the denominator have a pole
\begin{equation}
1 + \alpha - \beta \pm i \sqrt{C} = -n
\end{equation}
where $n=0,1,2,...$ is the overtone number. Given the time dependence of the scalar field, $\sim e^{-i \omega t}$, the mode is unstable (exponential growth) when $\omega_I > 0$ and stable (exponential decay) when $\omega_I < 0$. In the latter case the real part determines the frequency of the oscillation, $\omega_R/(2 \pi)$, while the inverse of $|\omega_I|$ determines the dumping time, $t_D^{-1}=|\omega_I|$. 

\begin{figure*}[ht]
\centering
\includegraphics[width=0.48\textwidth]{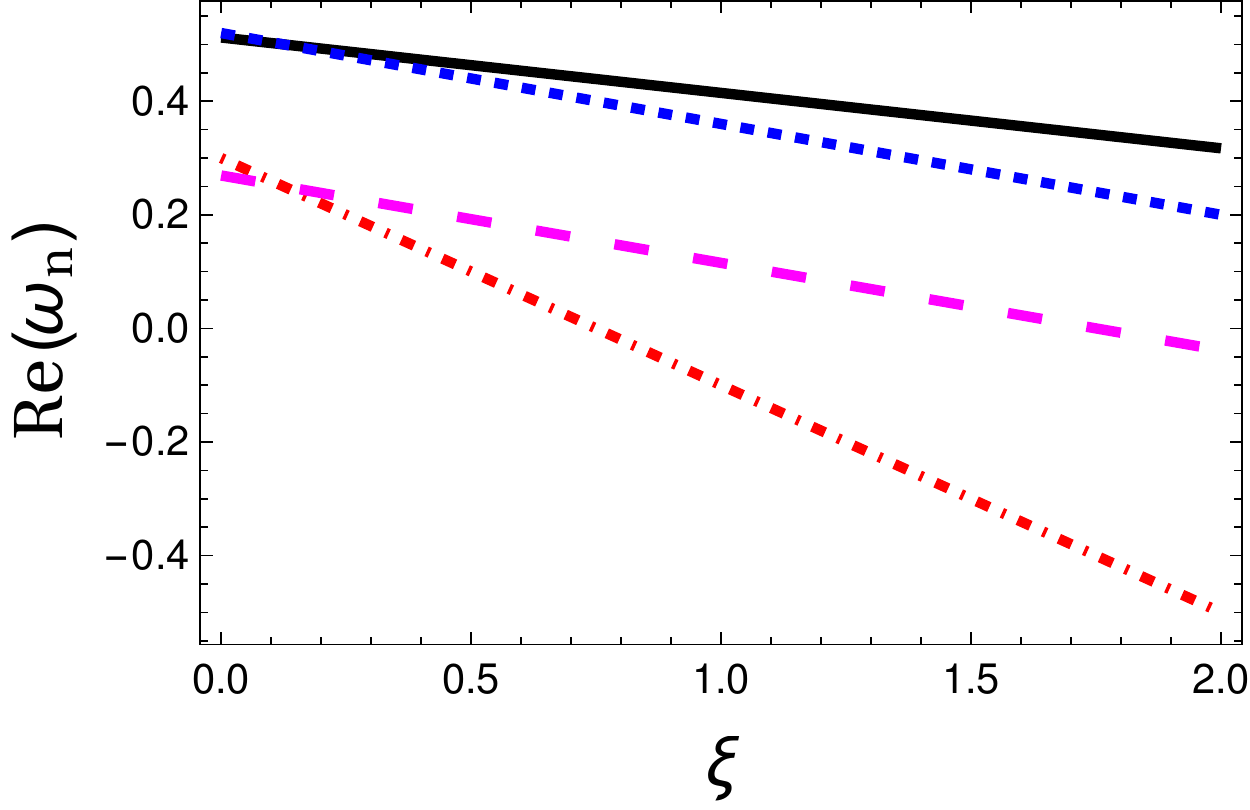}   
\ \ \
\includegraphics[width=0.48\textwidth]{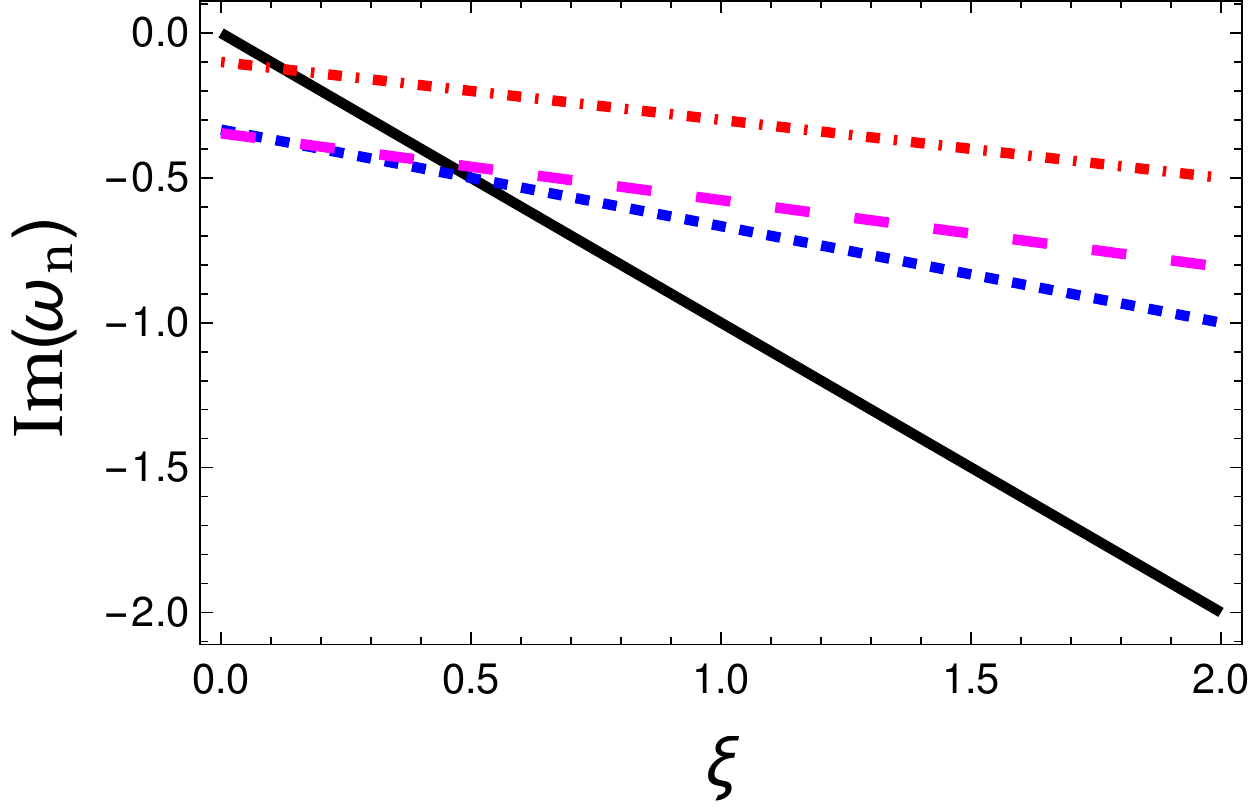}   

\caption{
Quasinormal frequencies $\omega_n$ vs non-minimal coupling. {\bf{Left panel:}} $\text{Re}(\omega_n)$ versus $\xi$ for four different cases: i)  $m=2=n$ (solid black line), ii) $m=2,n=1$ (blue short dashed line), iii) $m=1,n=0$ (dotted-dashed red line) and iv) $m=1=n$  (magenta long dashed line).  {\bf{Right panel:}} $\text{Im}(\omega_n)$ versus $\xi$ for four different cases: i)  $m=0=n$ (solid black line), ii) $m=0,n=1$ (blue short dashed line), iii) $m=1,n=0$ (dotted-dashed red line) and iv) $m=1=n$ (magenta long dashed line). In both figures we fix $r_H=2$ and $\eta = 0.25$.
}
\label{fig:frequencies}
\end{figure*}

The real part (left panel) and the imaginary part (right panel) of the quasinormal frequencies as a function of the nonminimal coupling are shown in Fig. \eqref{fig:frequencies}. We see that the slope decreases both with $m$ and with $n$. The real part vanishes for $m=0$, while for $m > 0$ it is initially positive, but eventually it becomes negative when $\xi$ becomes sufficiently large. Since it is a monotonically decreasing function of the nonminimal coupling, the frequency of the oscillation decreases with $\xi$. In particular, the period of the oscillation in the case of a non-vanishing $\xi$ is larger than the period corresponding to a canonical scalar field ($\xi=0$). 

Due to the emission of gravitational waves the spacetime undergoes dumbed oscillations, and this is encoded into the non-vanishing imaginary part. The latter is always negative, and therefore the scalar perturbations studied here are $\textit{stable}$. Since the imaginary part is a monotonically decreasing function of the nonminimal coupling, a certain mode for given $(m,n)$ decays faster as $\xi$ increases.

\section{Conclusion}

In the present article we have studied the propagation of a test massless nonminimally coupled scalar field into a fixed gravitational background of a cloud of strings, extending a previous work where the probe field was a scalar field with a canonical kinetic term minimally coupled to gravity. We have obtained the expression for the effective potential, and we have solved the full radial equation exactly in terms of hypergeometric functions. We thus have obtained exact analytical expressions for the reflection coefficient, the absorption cross section as well as the decay rate. Finally, applying the quasinormal boundary condition we have obtained the expression for the quasinormal spectrum. The impact of the nonminimal coupling has been investigated in detail. Our results show that Universality is not respected in general, and that scalar perturbations are stable.

\section*{Acknowlegements}

The work of A.R. was supported by the CONICYT-PCHA/ Doctorado Nacional/2015-21151658. G.P. thanks the Fun\-da\c c\~ao para a Ci\^encia e Tecnologia (FCT), Portugal, for the financial support to the Center for Astrophysics and Gravitation-CENTRA, Instituto Superior T\'ecnico, Universidade de Lisboa, through the Grant No. UID/FIS/ 00099/2013.

\end{document}